\documentclass[12pt]{article}
\usepackage{mathrsfs}
\usepackage{graphicx}
\usepackage{amsmath}
\usepackage{amssymb}
\usepackage{caption2}
\begin{document}
\newcommand  {\ba} {\begin{eqnarray}}
\newcommand  {\be} {\begin{equation}}
\newcommand  {\ea} {\end{eqnarray}}
\newcommand  {\ee} {\end{equation}}
\renewcommand{\thefootnote}{\fnsymbol{footnote}}
\renewcommand{\figurename}{Figure.}
\renewcommand{\captionlabeldelim}{.~}

\vspace*{1cm}
\begin{center}
 {\Large\textbf{Fermion Masses and Flavor Mixings from Family Symmetry $SU(3)$}}

\vspace{1cm}
 \textbf{Wei-Min Yang\footnote{E-mail address: wmyang@ustc.edu.cn}}

\vspace{0.3cm}
 \emph{Department of Modern Physics, University of Science and Technology of China, Hefei 230026, P. R. China}
\end{center}

\vspace{1cm}
 \noindent\textbf{Abstract}: We suggest a new particle model based on the symmetry group
 $SU(3)_{C}\otimes SU(2)_{L}\otimes SU(2)_{L'}\otimes SU(2)_{R}\otimes U(1)_{B\mbox{-}L}\otimes SU(3)_{F}\otimes U(1)_{N}$\,. The family symmetry and the high-energy left-handed and right-handed isospin subgroups are respectively broken by some flavon and Higgs fields one after another. At the low-energy scale the super-heavy fermions are all integrated out, the model finally leads to an effective theory with the standard model symmetry group. After the electroweak breaking all the fermion mass matrices are elegantly characterized only by six parameters. The model can perfectly fit and explain all the current experimental data about the fermion masses and flavor mixings, in particular, it finely predicts the first generation quark masses and the values of $\theta^{\,l}_{13}\,, \langle m_{\beta\beta}\rangle\,, J_{CP}^{\,l}$ in neutrino physics. The results are all promising to be tested in future experiments.

\vspace{1cm}

\vspace{0.3cm}

\newpage
 \noindent\textbf{I. Introduction}

\vspace{0.3cm}
 The precise tests for the electroweak scale physics have established plenty of knowledge about the elementary particles \cite{1}. The standard model (SM) has been evidenced to be indeed a very successful theory at the current energy scale \cite{2}. However, one of the SM defects is that too many parameters exist in the Yukawa sector. This leads that fermion
 masses and flavor mixings seem intricate and ruleless. The researches on this field always attract great attention in particle physics\cite{3}. In particular, during the past decade a series of new experiment results about $B$ physics and neutrino physics tell us a great deal of information about flavor physics \cite{4}. What deserves to be paid special attention are
 some facts as follows. The mass spectrum of quarks and charged leptons emerges a large hierarchy, which ranges from one MeV
 to a hundred GeV or so \cite{1}. The left-handed neutrinos have been verified to have non-zero but Sub-eV masses \cite{5}, nevertheless, that their nature is Majorana or Dirac particle has to be further identified by experiments such as $0\nu\beta\beta$ \cite{6}. On the other hand, the flavor mixing in the quark sector is distinctly different from that in the lepton sector. The former has small mixing angles and its mixing matrix is close to an unit matrix \cite{7}, whereas the latter has bi-large mixing angles and its mixing matrix is close to the tri-bimaximal mixing pattern \cite{8}. In the lepton mixing, it is yet in suspense whether $sin\theta_{13}$ is zero and the $CP$ violation vanishes or not \cite{9}. These impressive puzzles are always expected to be explained by new theories beyond the SM. The issues in the flavor physics implicate great significance. They are not only bound up with origin of matter in the universe \cite{10}, but also they are possibly in connection with the genesis of the matter-antimatter asymmetry and the original nature of the dark matter \cite{11}.

 Any new theory beyond the SM has to be confronted with the diverse intractable issues mentioned above, however, Some theoretical models have been proposed to solve them \cite{12}. For instance, the Froggatt-Nielsen mechanism with $U(1)$ family symmetry can account for mass hierarchy \cite{13}. The discrete family group $A_{4}$ can lead to the tri-bimaximal mixing structure of the lepton mixing matrix \cite{14a}. The non-Abelian continuous group $SU(3)$ is introduced to explain the neutrino mixing \cite{14b}. By means of the family group $SO(3)$ in \cite{15}, the model accommodates successfully the whole experimental data of quarks and leptons. In addition, some models of grand unification (GUT) based on $SO(10)$ symmetry group can also give some reasonable interpretation for fermion masses and flavor mixings \cite{16}. Although these models seem successful in explaining some flavor problems, it seems very difficult for them to solve all the flavor problems all together. It is especially hard for some models to keep to the principle of the smaller number of parameters. It remains to be a large challenge for theoretical particle physicists to uncover these mysteries of the flavor physics.

 In this works, we consider a new approach to solve the above problems and construct a model with fewer parameters. First of all, we believe that there is some inherence relations among all kinds of fermion mass and mixing parameters. The family symmetry $SU(3)_{F}$, which is a family symmetry group of the three generation fermions, is appropriate for seeking the relations. In addition, we introduce some super-heavy fermion and flavon fields which appear only at the high-energy scale. They have Yukawa couplings with the low-energy fermions of the SM. The different super-heavy fermions are distinguished by
 an appended Abelian group $U(1)_{N}$. On the other hand, the left-right symmetry group $SU(3)_{C}\otimes SU(2)_{L}\otimes SU(2)_{R}\otimes U(1)_{B\mbox{-}L}$ is theoretically well-motivated extension of the SM symmetry group \cite{17}. We extend
 the left-right symmetry group to include a new subgroup $SU(2)_{L'}$, which is a high-energy isospin symmetry group of the left-handed super-heavy fermions. The model goes through three steps of breakings. The family symmetry $SU(3)_{F}\otimes U(1)_{N}$ is firstly broken at the flavon dynamics scale. It is accomplished by means of the flavon fields developing the special vacuum structures. Secondly, the subgroups $SU(2)_{L'}\otimes SU(2)_{R}\otimes U(1)_{B\mbox{-}L}$ are broken at the middle energy scale. Lastly the electroweak symmetry is broken at the electroweak scale. The last two breakings are implemented respectively by the arranged high-energy and low-energy Higgs fields. After integrating all the super-heavy fermions out, the low-energy effective theory is elegantly obtained. All the fermion mass matrices are clearly given and characterized only by the six parameters. Finally, the theoretical structures of the model can naturally give rise to the correct fermion mass spectrum and flavor mixing angles. All the numerical results are very well in agreement with the current experimental data.

 The remainder of this paper is organized as follows. In Section II we outline the model. In Sec. III, the symmetry breaking procedure is introduced and the fermion mass matrices are discussed. In Sec. IV, we give the detailed numerical results about the fermion masses and flavor mixings. Sec. V is devoted to conclusions.

\vspace{1cm}
 \noindent\textbf{II. Model}

\vspace{0.3cm}
 We now outline the model. It is based on the symmetry group
 $SU(3)_{C}\otimes SU(2)_{L}\otimes SU(2)_{L'}\otimes SU(2)_{R}\otimes U(1)_{B\mbox{-}L}\otimes SU(3)_{F}\otimes U(1)_{N}$\,.
 Among them, $SU(3)_{C}$ and $SU(2)_{L}$ are the color and left-handed weak isospin subgroups of the SM at the low energy
 scale. The subgroups $SU(2)_{L'}\otimes SU(2)_{R}\otimes U(1)_{B\mbox{-}L}$ are the left-handed and right-handed isospin and $B\mbox{-}L$ symmetry group at the high energy scale. The family symmetry at the higher energy scale is characterized by the subgroups $SU(3)_{F}\otimes U(1)_{N}$. The particle contents and their quantum numbers under the model symmetry group are listed as follows. The low-energy matter fermion fields are
\begin{alignat}{1}
 & Q_{L}=\left(\begin{array}{c}u^{i}_{L}\\d^{i}_{L}\end{array}\right)_{\alpha}\sim(3,2,1,1,\frac{1}{3},3,0)\,,\hspace{0.4cm}
   Q_{R}=\left(\begin{array}{c}u^{i}_{R}\\d^{i}_{R}\end{array}\right)_{\alpha}\sim(3,1,1,2,\frac{1}{3},3,0)\,,\nonumber\\
 & L_{L}=\left(\begin{array}{c}\nu_{L}\\e_{L}\end{array}\right)_{\alpha}\sim(1,2,1,1,-1,3,0)\,,\hspace{0.3cm}
   L_{R}=\left(\begin{array}{c}\nu_{R}\\e_{R}\end{array}\right)_{\alpha}\sim(1,1,1,2,-1,3,0)\,,
\end{alignat}
 where the letters $\alpha$ and $i$ are respectively family and color indices. The left-handed and right-handed fermion fields are in the respective doublet representations under the subgroups $SU(2)_{L}$ and $SU(2)_{R}$\,. The quarks and leptons are respectively triplet and singlet under the color subgroup $SU(3)_{C}$\,, and also they have the different $B\mbox{-}L$ quantum numbers. In any case, the three generation fermions are in $\mathbf{3}$ representation of the family subgroup $SU(3)_{F}$\,, moreover, they don't have any charges of the subgroup $U(1)_{N}$\,.

 We introduce some super-heavy fermion fields as follows, all of which are possessed the super-heavy masses and appear only in the very high energy circumstances. The super-heavy quarks are
\begin{alignat}{1}
 &\eta_{0L}\sim(3,1,2,1,\frac{1}{3},3,-1)\,,\hspace{0.3cm}\eta_{0R}\sim(3,1,1,2,\frac{1}{3},3,-6)\,,\nonumber\\
 &\eta_{1L}\sim(3,1,2,1,\frac{1}{3},3,0)\,,\hspace{0.6cm}\eta_{1R}\sim(3,1,1,2,\frac{1}{3},3,-1)\,,\nonumber\\
 &\eta_{2L}\sim(3,1,2,1,\frac{1}{3},3,1)\,,\hspace{0.6cm}\eta_{2R}\sim(3,1,1,2,\frac{1}{3},3,-2)\,,\nonumber\\
 &\eta_{3L}\sim(3,1,2,1,\frac{1}{3},3,2)\,,\hspace{0.6cm}\eta_{3R}\sim(3,1,1,2,\frac{1}{3},3,-4)\,,\nonumber\\
 &\eta_{4L}\sim(3,1,2,1,\frac{1}{3},3,4)\,,\hspace{0.6cm}\eta_{4R}\sim(3,1,1,2,\frac{1}{3},3,-3)\,,\nonumber\\
 &\eta_{5L}\sim(3,1,2,1,\frac{1}{3},3,3)\,,\hspace{0.6cm}\eta_{5R}\sim(3,1,1,2,\frac{1}{3},3,-5)\,,
\end{alignat}
 and the super-heavy leptons are
\begin{alignat}{1}
 &\xi_{1L}\sim(1,1,2,1,-1,3,0)\,,\hspace{0.6cm}\xi_{1R}\sim(1,1,1,2,-1,3,-1)\,,\nonumber\\
 &\xi_{2L}\sim(1,1,2,1,-1,3,1)\,,\hspace{0.6cm}\xi_{2R}\sim(1,1,1,2,-1,3,-2)\,,\nonumber\\
 &\xi_{3L}\sim(1,1,2,1,-1,3,-2)\,,\hspace{0.3cm}\xi_{3R}\sim(1,1,1,2,-1,3,-4)\,,\nonumber\\
 &\xi_{4L}\sim(1,1,2,1,-1,3,2)\,,\hspace{0.6cm}\xi_{4R}\sim(1,1,1,2,-1,3,-3)\,,\nonumber\\
 &\xi_{5L}\sim(1,1,2,1,-1,3,5)\,,\hspace{0.6cm}\xi_{5R}\sim(1,1,1,2,-1,3,-5)\,.
\end{alignat}
 In comparison with the super-heavy quark fields, the super-heavy lepton fields have not the $\xi_{0L}$ and $\xi_{0R}$ terms. All the super-heavy fermions are singlets under $SU(2)_{L}$. Their left-handed fields are doublets under the left-handed isospin group $SU(2)_{L'}$, meanwhile, the right-handed fields are doublets under the right-handed isospin group $SU(2)_{R}$\,. Each super-heavy fermion fields has their respective quantum numbers of $U(1)_{N}$\,. However, their color, $B\mbox{-}L$ and family quantum numbers are the same as those of the low-energy fermions.

 The Higgs fields in the model include
\begin{alignat}{1}
 &H=\left(\begin{array}{cc}H^{0}_{2}&H^{+}_{1}\\H^{-}_{2}&H^{0}_{1}\end{array}\right)\sim(1,2,1,2,0,1,0)\,,\nonumber\\
 &\Theta=\left(\begin{array}{c}\Theta^{0}\\\Theta^{-}\end{array}\right)\sim(1,1,1,2,-1,1,0)\,,\nonumber\\
 &\Omega_{1}\sim(1,1,2,2,0,1,-1)\,,\hspace{0.3cm}\Omega_{2}\sim(1,1,2,2,0,1,-3)\,,\nonumber\\
 &\Omega_{3}\sim(1,1,2,2,0,1,6)\,,\hspace{0.6cm}\Omega_{4}\sim(1,1,2,2,0,1,0)\,,\nonumber\\
 &\Omega_{5}\sim(1,1,2,2,0,1,2)\,.
\end{alignat}
 In addition, we also introduce $\widetilde{\Omega_{i}}=\tau_{2}\Omega_{i}^{*}\tau_{2}$ $(i=1,2,\cdots,5)$. Here and thereinafter $\tau_{1}, \tau_{2}, \tau_{3}$ are Pauli matrices. The Higgs field $H$ is responsible for the electroweak
 symmetry breaking at the low-energy scale. The Higgs fields $\Theta$ and $\Omega_{i}$ play roles in breaking of the
 high-energy symmetry subgroups $SU(2)_{L'}\otimes SU(2)_{R}\otimes U(1)_{B\mbox{-}L}$.

 We finally introduce the super-heavy scalar flavon fields
\begin{alignat}{1}
 & \Phi_{0}\sim(1,1,1,1,0,1,1)\,,\hspace{0.4cm}F_{1}\sim(1,1,1,1,0,8,-3)\,,\hspace{0.3cm}F_{2}\sim(1,1,1,1,0,8,2)\,,\nonumber\\
 & F_{3}\sim(1,1,1,1,0,8,-4)\,,\hspace{0.1cm} F_{4}\sim(1,1,1,1,0,8,4)\,,\hspace{0.6cm}F_{5}\sim(1,1,1,1,0,8,3)\,,\nonumber\\
 & F_{6}\sim(1,1,1,1,0,\overline{6},0)\,,\hspace{0.4cm} F_{7}\sim(1,1,1,1,0,\overline{6},-1)\,,\hspace{0.3cm}
   F_{8}\sim(1,1,1,1,0,\overline{6},-2)\,.
\end{alignat}
 Under the family subgroup $SU(3)_{F}$, $\Phi_{0}$ is a real singlet representation, and $F_{1}, \cdots, F_{5}$ are all hermitian octet representations, and $F_{6}, \cdots, F_{8}$ are all complex symmetric sextet representations. In addition,
 these flavon fields have different charges of $U(1)_{N}$\,. They are responsible for the family symmetry breaking.

 Under the model symmetry group, the gauge invariant Yukawa couplings in the quark sector are such as
 \begin{alignat}{1}
 \mathscr{L}_{q}=
 &\:y_{q}\left[\frac{\Phi_{0}^{6}}{\Lambda_{F}^{6}}\,\overline{Q_{L}}H\eta_{0R}+\overline{\eta_{0L}}\,\Omega_{1}Q_{R}
   +\frac{\Phi_{0}^{8}}{\Lambda_{F}^{8}}\,\overline{\eta_{0L}}\,\Omega_{2}\eta_{0R}\right.\nonumber\\
 &+\frac{\Phi_{0}}{\Lambda_{F}}\,\overline{Q_{L}}H\eta_{1R}+\frac{F_{1}}{\Lambda_{F}}\,\overline{\eta_{1L}}\,\widetilde{\Omega_{2}}Q_{R}
   +\frac{\Phi_{0}^{4}}{\Lambda_{F}^{4}}\,\overline{\eta_{1L}}\,\Omega_{2}\eta_{1R}\nonumber\\
 &+\frac{\Phi_{0}^{2}}{\Lambda_{F}^{2}}\,\overline{Q_{L}}H\eta_{2R}+\frac{F_{2}}{\Lambda_{F}}\,\overline{\eta_{2L}}\,\Omega_{1}Q_{R}
   +\frac{\Phi_{0}^{4}}{\Lambda_{F}^{4}}\,\overline{\eta_{2L}}\,\Omega_{1}\eta_{2R}\nonumber\\
 &+\frac{\Phi_{0}^{4}}{\Lambda_{F}^{4}}\,\overline{Q_{L}}H\eta_{3R}+\frac{F_{3}}{\Lambda_{F}}\,\overline{\eta_{3L}}\,\Omega_{3}Q_{R}
   +\frac{\Phi_{0}^{5}}{\Lambda_{F}^{5}}\,\overline{\eta_{3L}}\,\widetilde{\Omega_{1}}\eta_{3R}\nonumber\\
 &+\frac{\Phi_{0}^{3}}{\Lambda_{F}^{3}}\,\overline{Q_{L}}H\eta_{4R}+\frac{F_{4}}{\Lambda_{F}}\,\overline{\eta_{4L}}\,\Omega_{4}Q_{R}
   +\frac{\Phi_{0}^{6}}{\Lambda_{F}^{6}}\,\overline{\eta_{4L}}\,\widetilde{\Omega_{1}}\eta_{4R}\nonumber\\
 &\left.+\frac{\Phi_{0}^{5}}{\Lambda_{F}^{5}}\,\overline{Q_{L}}H\eta_{5R}+\frac{F_{5}}{\Lambda_{F}}\,\overline{\eta_{5L}}\,\widetilde{\Omega_{4}}Q_{R}
   +\frac{\Phi_{0}^{7}}{\Lambda_{F}^{7}}\,\overline{\eta_{5L}}\,\widetilde{\Omega_{1}}\eta_{5R}\right]\nonumber\\
 & +h.c.\,,
\end{alignat}
 where $y_{q}$ is the uniform and only Yukawa coupling coefficient in the quark sector and it should be $\thicksim \mathscr{O}(1)$, in addition, $\Lambda_{F}$ is a dynamics scale of the family symmetry. Below the scale $\Lambda_{F}$\,,
 all the flavon fields develop the vacuum states, and consequently the family symmetry is broken. The Yukawa couplings in
 the lepton sector are similarly written as
\begin{alignat}{1}
 \mathscr{L}_{l}=
 &\:y_{l}\left[\frac{\Phi_{0}}{\Lambda_{F}}\,\overline{L_{L}}H\xi_{1R}+\frac{F_{1}}{\Lambda_{F}}\,\overline{\xi_{1L}}\,\widetilde{\Omega_{2}}L_{R}
   +\frac{\Phi_{0}^{4}}{\Lambda_{F}^{4}}\,\overline{\xi_{1L}}\,\Omega_{2}\xi_{1R}\right.\nonumber\\
 &+\frac{\Phi_{0}^{2}}{\Lambda_{F}^{2}}\,\overline{L_{L}}H\xi_{2R}+\frac{F_{2}}{\Lambda_{F}}\,\overline{\xi_{2L}}\,\Omega_{1}L_{R}
   +\frac{\Phi_{0}^{4}}{\Lambda_{F}^{4}}\,\overline{\xi_{2L}}\,\Omega_{1}\xi_{2R}\nonumber\\
 &+\frac{\Phi_{0}^{4}}{\Lambda_{F}^{4}}\,\overline{L_{L}}H\xi_{3R}+\frac{F_{3}}{\Lambda_{F}}\,\overline{\xi_{3L}}\,\Omega_{5}L_{R}
   +\frac{\Phi_{0}^{5}}{\Lambda_{F}^{5}}\,\overline{\xi_{3L}}\,\Omega_{2}\xi_{3R}\nonumber\\
 &+\frac{\Phi_{0}^{3}}{\Lambda_{F}^{3}}\,\overline{L_{L}}H\xi_{4R}+\frac{F_{4}}{\Lambda_{F}}\,\overline{\xi_{4L}}\,\widetilde{\Omega_{5}}L_{R}
   +\frac{\Phi_{0}^{6}}{\Lambda_{F}^{6}}\,\overline{\xi_{4L}}\,\Omega_{1}\xi_{4R}\nonumber\\
 &\left.+\frac{\Phi_{0}^{5}}{\Lambda_{F}^{5}}\,\overline{L_{L}}H\xi_{5R}+\frac{F_{5}}{\Lambda_{F}}\,\overline{\xi_{5L}}\,\Omega_{5}L_{R}
   +\frac{\Phi_{0}^{7}}{\Lambda_{F}^{7}}\,\overline{\xi_{5L}}\,\widetilde{\Omega_{2}}\xi_{5R}\right]\nonumber\\
 & +h.c.\,,
\end{alignat}
 likewise, $y_{l}$ is the uniform and only Yukawa coupling coefficient in the lepton sector and it is $\thicksim \mathscr{O}(1)$. In comparison with the quark sector, the lepton sector is short of the terms related to $\overline{\xi_{0L}}$ and $\xi_{0R}$\,, in addition, these Higgs fields coupled with $\overline{\xi_{3L}},\overline{\xi_{4L}},\overline{\xi_{5L}}$ are different from those in the quark sector. These differences play key roles in generating the distinct masses and mixings
 of the quarks and leptons. Finally, the right-handed leptons have the characteristic Majorana-type couplings
\ba
 \mathscr{L}_{RM}
  =y_{R}L_{R}^{T}\,\Theta^{*}\!\left[\frac{F_{6}}{\Lambda_{F}}+\frac{\Phi_{0}}{\Lambda_{F}}\frac{F_{7}}{\Lambda_{F}}
   +\frac{\Phi_{0}^{2}}{\Lambda_{F}^{2}}\frac{F_{8}}{\Lambda_{F}}\right]\!\frac{\Theta^{\dagger}}{\Lambda_{F}}L_{R}\,,
\ea
 where $y_{R}$ is also a coupling coefficient, and it is $\thicksim \mathscr{O}(1)$. Here we have left out the charge conjugation matrix $C$ sandwiched between two spinor fields, hereinafter so does. The Majorana couplings will generate
 Majorana masses of the right-handed neutrinos after the doublet Higgs $\Theta$ develops the vacuum expectation value (VEV).

\vspace{1cm}
 \noindent\textbf{III. Symmetry Breakings and Fermion Mass Matrices}

\vspace{0.3cm}
 The model symmetry breakings go through three stages. The first step of the breaking chain is that the subgroups $SU(3)_{F}\otimes U(1)_{N}$ break to nothing, namely which means that the family symmetry vanishes. This is accomplished by
 the flavon fields $\Phi_{0}, F_{1}, \cdots, F_{8}$ developing VEVs which are slightly larger than the scale $\Lambda_{F}$\,. The specific vacuum structures of the flavon fields are as follows
\begin{alignat}{1}
 & \frac{\langle\Phi_{0}\rangle}{\Lambda_{F}}=\frac{1}{\varepsilon_{0}}\,,\hspace{0.5cm}
   \frac{\langle F_{1}\rangle}{\langle\Phi_{0}\rangle}
         =\left(\begin{array}{ccc}0&-3&-3\\-3&0&-3\\-3&-3&0\\\end{array}\right)\!,\hspace{0.5cm}
   \frac{\langle F_{2}\rangle}{\langle\Phi_{0}\rangle}
         =\left(\begin{array}{ccc}0&0&0\\0&1&-1\\0&-1&1\\\end{array}\right)\!,\nonumber\\
 & \frac{\langle F_{3}\rangle}{\langle\Phi_{0}\rangle}
         =\left(\begin{array}{ccc}0&0&0\\0&0&0\\0&0&\frac{2}{3}\\\end{array}\right)\!,\hspace{0.1cm}
   \frac{\langle F_{4}\rangle}{\langle\Phi_{0}\rangle}
         =\left(\begin{array}{ccc}0&0&3+i\\0&0&0\\3-i&0&0\\\end{array}\right)\!,\hspace{0.1cm}
   \frac{\langle F_{5}\rangle}{\langle\Phi_{0}\rangle}
         =\left(\begin{array}{ccc}0&\frac{-1}{4}&\frac{-i}{4\sqrt{3}}\\\frac{-1}{4}&0&\frac{1}{2}\\\frac{i}{4\sqrt{3}}&\frac{1}{2}&0\\\end{array}\right)\!,\nonumber\\
 & \frac{\langle F_{6}\rangle}{\langle\Phi_{0}\rangle}
         =\left(\begin{array}{ccc}2&0&0\\0&2&0\\0&0&2\\\end{array}\right)\!,\hspace{0.1cm}
   \frac{\langle F_{7}\rangle}{\langle\Phi_{0}\rangle}
         =\left(\begin{array}{ccc}-\frac{3}{2}&1&1\\1&-\frac{3}{2}&1\\1&1&-\frac{3}{2}\\\end{array}\right)\!,\hspace{0.1cm}
   \frac{\langle F_{8}\rangle}{\langle\Phi_{0}\rangle}=\left(\begin{array}{ccc}0&0&0\\0&1&-1\\0&-1&1\\\end{array}\right)\!.
\end{alignat}
 All of the matrix elements, which are almost $\thicksim \mathscr{O}(1)$, are determined by the vacuum structures. The only one undetermined value is $\varepsilon_{0}$\,, which is the ratio of the family symmetry scale to the singlet flavon field VEV. We consider that $\langle\Phi_{0}\rangle$ is two orders of magnitude higher than $\Lambda_{F}$\,, thus $\varepsilon_{0}$ should be a small quantity about $10^{-2}\thicksim 10^{-3}$. It can be seen from (9) that the breakings of $F_{1},F_{6},F_{7}$ occur along direction of the subgroup $S_{3}$ in the family space, which is a permutation group among three generation fermions. The $F_{2}$ and $F_{8}$ breakings are in the direction of the subgroup $S_{2}$ ($2 \leftrightarrow 3$ permutation), while the $F_{3}$ breaking is in the direction of the subgroup $S_{2}'$ ($1 \leftrightarrow 2$ permutation). The breakings of $F_{4}$ and $ F_{5}$ result in the family symmetry being lost eventually, moreover, their imaginary elements are also sources of the $C$ and $CP$ violation. Of course, these vacuum structures should essentially be determined by the self-interaction potential of every flavon field. Here we do not go into detailed discussion about them, but accept the specific breaking mode since it turns out to be a great success in fitting experimental data later.

 The second step of the breaking chain is that the subgroups $SU(2)_{L'}\otimes SU(2)_{R}\otimes U(1)_{B\mbox{-}L}$ break to $U(1)_{Y}$\,. This is achieved by the high-energy Higgs fields $\Omega_{1},\cdots,\Omega_{5}$ and $\Theta$ developing VEVs at the scale $\Lambda_{R}$ as follows
\begin{alignat}{1}
 & \frac{\langle\Omega_{1}\rangle}{\Lambda_{R}}=\left(\begin{array}{cc}-2&0\\0&2\\\end{array}\right),\hspace{0.2cm}
   \frac{\langle\Omega_{2}\rangle}{\Lambda_{R}}=\left(\begin{array}{cc}-6&0\\0&-6\\\end{array}\right),\hspace{0.2cm}
   \frac{\langle\Omega_{3}\rangle}{\Lambda_{R}}=\left(\begin{array}{cc}3&0\\0&1\\\end{array}\right),\nonumber\\
 & \frac{\langle\Omega_{4}\rangle}{\Lambda_{R}}=\left(\begin{array}{cc}2&0\\0&0\\\end{array}\right),\hspace{0.5cm}
   \frac{\langle\Omega_{5}\rangle}{\Lambda_{R}}=\left(\begin{array}{cc}0&0\\0&-1\\\end{array}\right),\hspace{0.6cm}
   \frac{\langle\Theta\rangle}{\Lambda_{R}}=\left(\begin{array}{c}1\\0\end{array}\right).
\end{alignat}
 The breaking scale $\Lambda_{R}$ should be an intermediate value between the family breaking scale $\Lambda_{F}$ and the electroweak breaking scale. Therefore it is far smaller than the scale $\Lambda_{F}$ and also far larger than the electroweak scale. The Higgs vacuum structures are of course determined by their self-interaction potential. The new subgroup $U(1)_{Y}$
 is a linear combination from the original subgroups $U(1)_{I^{L'}_{3}}$, $U(1)_{I^{R}_{3}}$ and $U(1)_{B\mbox{-}L}$\,. Their charge quantum numbers are related by the formula
\ba
 \frac{Y}{2}=I^{L'}_{3}+I^{R}_{3}+\frac{B-L}{2}\,.
\ea

 Below the scale $\Lambda_{R}$\,, the model remaining symmetry is namely the SM symmetry group. Now it can be seen from the Lagrangian (6), (7), (8) that all of the super-heavy fermions achieve Dirac masses, and that the right-handed neutrinos generate Majorana masses. Because all of the super-heavy fermion masses are far greater than their couplings with the SM fermions, at the low energy all of them are actually decoupling. After all the super-heavy fermions are integrated out
 from the original Lagrangian, then an effective Yukawa Lagrangian is derived as
\begin{alignat}{1}
 \mathscr{L}_{Yukawa}=&\:\overline{Q_{L}}\,H_{2}\,Y_{u}\,u_{R}+\overline{Q_{L}}\,H_{1}\,Y_{d}\,d_{R}
                        +\overline{L_{L}}\,H_{1}\,Y_{e}\,e_{R}\nonumber\\
  & +\overline{L_{L}}\,H_{2}\,Y_{D}\,\nu_{R}-\frac{1}{2}\,\nu_{R}^{T}\,M_{R}\,\nu_{R}+h.c.
\end{alignat}
 with Yukawa coupling matrices and Majorana mass matrix of the right-handed neutrinos
\begin{alignat}{1}
 & Y_{u}=-y_{q}\left(\frac{1}{3}\,\varepsilon_{0}^{2}\,I+\varepsilon_{0}^{2}\,\widetilde{F_{1}}
         +\varepsilon_{0}\widetilde{F_{2}}+\frac{3}{2}\,\widetilde{F_{3}}+\varepsilon_{0}^{2}\,\widetilde{F_{4}}\right),\nonumber\\
 & Y_{d}=-y_{q}\left(-\frac{1}{3}\,\varepsilon_{0}^{2}\,I+\varepsilon_{0}^{2}\,\widetilde{F_{1}}
         +\varepsilon_{0}\widetilde{F_{2}}-\frac{1}{2}\,\widetilde{F_{3}}-\varepsilon_{0}\widetilde{F_{5}}\right),\nonumber\\
 & Y_{D}=-y_{l}\left(\varepsilon_{0}^{2}\,\widetilde{F_{1}}+\varepsilon_{0}\widetilde{F_{2}}
         +\frac{1}{2}\,\varepsilon_{0}^{2}\,\widetilde{F_{4}}\right),\nonumber\\
 & Y_{e}=-y_{l}\left(\varepsilon_{0}^{2}\,\widetilde{F_{1}}+\varepsilon_{0}\widetilde{F_{2}}
         +\frac{1}{6}\,\widetilde{F_{3}}+\frac{1}{6}\,\varepsilon_{0}\widetilde{F_{5}}\right),\nonumber\\
 & M_{R}=-\frac{2\,y_{R}\,\Lambda_{R}^{2}}{\varepsilon_{0}^{3}\,\Lambda_{F}}\left(2\,\varepsilon_{0}^{2}\,I
         +\varepsilon_{0}\widetilde{F_{7}}+\widetilde{F_{2}}\right),
\end{alignat}
 where $I$ is a $3\times3$ unit matrix and each $\widetilde{F_{k}}$ is respectively the corresponding matrix in (9) (note $\widetilde{F_{8}}=\widetilde{F_{2}}$)\,. This low-energy effective theory, which is valid until the scale $\Lambda_{R}$\,,
 has two Higgs doublets. They are from the $H$ decomposition under the subgroups $SU(2)_{L}\otimes U(1)_{Y}$\,. In addition, there are also three right-handed Majorana neutrino singlets. Their masses are about the magnitude of $\Lambda_{R}^{2}/\varepsilon_{0}^{2}\Lambda_{F}$\,, which should be slightly smaller than the scale $\Lambda_{R}$\,. It is
 very clear from (13) that there are indeed some inherence relations among this set of Yukawa coupling matrices. They have
 three notable characteristics. First, every Yukawa coupling matrix is expanded by a power series of $\varepsilon_{0}$\,,
 thus it's elements show themselves large hierarchy. The $\widetilde{F_{3}}$ and $\widetilde{F_{2}}$ terms, namely the $\varepsilon_{0}^{0}$ and $\varepsilon_{0}^{1}$ terms, respectively dominate the third and second generation fermion masses. The rest of terms make main contributions to the first generation fermion mass. Second, the structure features of the Yukawa coupling matrices lead that the transformation matrices diagonalizing $Y_{u}$, $Y_{d}$ and $Y_{e}$ are all close to unit matrices. By contrast, $Y_{D}$ has no the $\widetilde{F_{3}}$ term, so the transformation matrix diagonalizing it is approximately the tri-bimaximal mixing pattern. This is the principal source of generating distinct flavor mixings for quarks and leptons. Third, in all there are only four independent parameters $\varepsilon_{0}$, $y_{q}$, $y_{l}$ and $y_{R}\Lambda_{R}^{2}/\Lambda_{F}$ in the effective theory. Among them, $\varepsilon_{0}$ is the only one parameter influencing the flavor mixing matrices except for those terms related to the $I$ matrix. Other parameters are only some product factors in the matrices of (13), so they have no effect on the flavor mixings.

 The last step of the breaking chain is that the subgroups $SU(2)_{L}\otimes U(1)_{Y}$ break to $U(1)_{em}$\,, namely electroweak symmetry breaking. It is implemented by the low-energy Higgs fields $H_{1}$ and $H_{2}$ developing VEVs at
 the electroweak scale $\Lambda_{L}$ as follows
\ba
 \frac{\langle H_{1}\rangle}{\Lambda_{L}}=\left(\begin{array}{c}0\\cos\beta\end{array}\right),\hspace{0.5cm}
 \frac{\langle H_{2}\rangle}{\Lambda_{L}}=\left(\begin{array}{c}sin\beta\\0\end{array}\right),
\ea
 where $tan\beta$ is the ratio of the up-type VEV to the down-type VEV. The electroweak breaking gives rise to Dirac masses of all the quarks and leptons, furthermore, the tiny Majorana masses of the left-handed neutrinos are generated by the see-saw mechanism since $M_{R}\gg \Lambda_{L}$ \cite{18}. The whole fermion mass terms are eventually written as
\begin{alignat}{1}
 -\mathscr{L}_{mass}=&\:\overline{u_{L}}\,M_{u}\,u_{R}+\overline{d_{L}}\,M_{d}\,d_{R}+\overline{e_{L}}\,M_{e}\,e_{R}\nonumber\\
                     & +\frac{1}{2}\,\overline{\nu_{L}}\,M_{\nu}\,\overline{\nu_{L}}^{T}
                       +\frac{1}{2}\,\nu_{R}^{T}\,M_{R}\,\nu_{R}+h.c.
\end{alignat}
 with the mass matrices
\begin{alignat}{1}
 & M_{u}=-\Lambda_{L}sin\beta\,Y_{u}\,,\hspace{0.6cm} M_{d}=-\Lambda_{L}cos\beta\,Y_{d}\,,\nonumber\\
 & M_{D}=-\Lambda_{L}sin\beta\,Y_{D}\,,\hspace{0.5cm} M_{e}=-\Lambda_{L}cos\beta\,Y_{e}\,,\nonumber\\
 & M_{\nu}=-M_{D}M_{R}^{-1}M_{D}^{T}\,.
\end{alignat}
 Two new parameters $\Lambda_{L}$ and $tan\beta$ are now added into the model besides the foregoing four parameters. It can be seen from (13) and (16) that $\Lambda_{L}$ dominates mass scales of the quarks and the charged leptons, while $tan\beta$ is responsible for mass splits of the up-type and down-type fermions. Anyway, they have no influence on the flavor mixings, the mass hierarchy and the flavor mixings are still controlled only by $\varepsilon_{0}$\,. In a word, these mass matrices properly embody all the information that is about fermion mass hierarchy, flavor mixing and the $CP$ violation.

 In virtue of the model's intrinsic characteristics, the Dirac-type mass matrices are all hermitian and the Majorana-type mass matrices are all complex symmetry. All of fermion mass eigenvalues are therefore solved by diagonalizing the mass matrices as follows
\begin{alignat}{1}
 & U_{u}^{\dagger}\,M_{u}\,U_{u}=\mathrm{diag}\left(m_{u},m_{c},m_{t}\right),\hspace{0.4cm}
   U_{d}^{\dagger}\,M_{d}\,U_{d}=\mathrm{diag}\left(m_{d},m_{s},m_{b}\right),\nonumber\\
 & U_{e}^{\dagger}\,M_{e}\,U_{e}=\mathrm{diag}\left(m_{e},m_{\mu},m_{\tau}\right),\hspace{0.4cm}
   U_{\nu}^{\dagger}\,M_{\nu}\,U_{\nu}^{*}=\mathrm{diag}\left(m_{1},m_{2},m_{3}\right),\nonumber\\
 & U_{R}^{T}\,M_{R}\,U_{R}=\mathrm{diag}\left(M_{1},M_{2},M_{3}\right).
\end{alignat}
 It can easily be calculated from(9), (13) and (16) that since the $\widetilde{F_{3}}$ and $\widetilde{F_{2}}$ terms are respectively the leading and next-to-leading terms in the mass matrices, the second and third generation quark and charged lepton masses have the approximate solutions such as
 \begin{alignat}{1}
 & m_{c}\approx y_{q}\,\Lambda_{L}sin\beta(\varepsilon_{0}-\varepsilon_{0}^{2})\,,\hspace{0.7cm}
   m_{t}\approx y_{q}\,\Lambda_{L}sin\beta(\varepsilon_{0}+1)\,,\nonumber\\
 & m_{s}\approx y_{q}\,\Lambda_{L}cos\beta(\varepsilon_{0}+3\,\varepsilon_{0}^{2})\,,\hspace{0.5cm}
   m_{b}\approx y_{q}\,\Lambda_{L}cos\beta(\varepsilon_{0}-\frac{1}{3})\,,\nonumber\\
 & m_{\mu}\approx y_{l}\,\Lambda_{L}cos\beta(\varepsilon_{0}-9\,\varepsilon_{0}^{2})\,,\hspace{0.5cm}
   m_{\tau}\approx y_{l}\,\Lambda_{L}cos\beta(\varepsilon_{0}+\frac{1}{9})\,.
\end{alignat}
 In the leading approximation, it can be seen further that there are the mass relations
\ba
 \frac{m_{c}}{m_{t}}\approx \varepsilon_{0}\,,\hspace{0.6cm}
 \frac{m_{s}}{m_{b}}\approx -3\,\varepsilon_{0}\,,\hspace{0.6cm}
 \frac{m_{\mu}}{m_{\tau}}\approx 9\,\varepsilon_{0}\,.
\ea
 However, the first generation quark and charged lepton masses have no such simple analytic expressions about their approximate solutions since they depend on all the terms in the mass matrices. Finally, the flavor mixing matrices for the quarks and leptons are respectively given by \cite{19}
 \ba U_{u}^{\dagger}\,U_{d}=U_{CKM}\,,\hspace{0.5cm}
     U_{e}^{\dagger}\,U_{\nu}=U_{PMNS}\:\mathrm{diag}\left(e^{i\beta_{1}},e^{i\beta_{2}},1\right),
 \ea
 where $\beta_{1}$, $\beta_{2}$ are two Majorana phases in the lepton mixing matrix. The mixing angles and $CP$-violating phases in the unitary matrices $U_{CKM}$ and $U_{PMNS}$ are worked out by the standard parameterization in particle data group \cite{1}.

\vspace{1cm}
 \noindent\textbf{IV. Numerical Results}

\vspace{0.3cm}
 In this section, we present numerical results of our model. As is noted earlier, the model totally involves six independent parameters, namely one ratio $\varepsilon_{0}$\,, and two Yukawa coefficients $y_{q}$ and $y_{l}$\,, and two electroweak breaking parameters $\Lambda_{L}$ and $tan\beta$\,, and one combined parameter $y_{R}\Lambda_{R}^{2}/\Lambda_{F}$. Once this set of parameters are chosen as the input values, we can calculate the various output values of the fermion masses and flavor mixings by the foregoing results. Of course, all the output results can be compared with the current and future experimental data.

 First of all, the electroweak breaking scales $\Lambda_{L}$ is only one product factor in the mass matrices. It is not truly a free parameters in Yukawa sector but rather should be determined in the gauge sector. The accurate measures about weak gauge boson masses and gauge coupling constant have given $\Lambda_{L}=174$ GeV. Secondly, the other five parameters are verily relevant to the fermion flavor issues. They can only be determined by fitting the experimental data in Yukawa sector and neutrino physics. We choose a set of the following values as input
\begin{alignat}{1}
 & \varepsilon_{0}=0.00748\,,\hspace{0.5cm} y_{q}=0.98\,,\hspace{0.5cm} y_{l}=1.139\,,\nonumber\\
 & tan\beta=13.24\,,\hspace{0.4cm} \frac{y_{R}\Lambda_{R}^{2}}{\Lambda_{F}}=1.91\times 10^{4}\:\mathrm{GeV}\,.
\end{alignat}
 All the values are completely consistent with the prior estimate. However, both of the $\Lambda_{R}$ and $\Lambda_{F}$ values are actually unknown. We only estimate their combined value $\Lambda_{R}^{2}/\Lambda_{F}\thicksim 10^{4}$ GeV. As a result, $\Lambda_{R}$ will be $10^{10}$ GeV if $\Lambda_{F}$ is at the GUT scale $10^{16}$ GeV. Finally, a variety of the numerical results predicted by the model are in detail listed the following.

 For the quark sector, all of mass eigenvalues and mixing angles are (mass in GeV unit)
\begin{alignat}{1}
 & m_{u}=0.00253\,,\hspace{0.6cm} m_{c}=1.27\,,\hspace{0.7cm} m_{t}=171.3\,;\nonumber\\
 & m_{d}=0.00475\,,\hspace{0.6cm} m_{s}=0.105\,,\hspace{0.5cm} m_{b}=4.19\,; \nonumber\\
 & s^{\,q}_{12}=0.2254\,,\hspace{0.2cm} s^{\,q}_{23}=0.0417\,,\hspace{0.2cm} s^{\,q}_{13}=0.00360\,,\hspace{0.2cm}
   \delta^{\,q}=0.379\,\pi\approx68.2^{\circ}\,,
\end{alignat}
 where $s_{\alpha\beta}=sin\theta_{\alpha\beta}$\,. Moreover, the Jarlskog invariant measuring the $CP$ violation is calculated to
\ba
 J_{CP}^{\,q}\approx3.06\times10^{-5}\,.
\ea
 The above results are very well in agreement with the current measures about the quark masses and mixing as well as the $CP$ violation \cite{1}. In particular, the first generation quark masses are finely forecast although their precise values have not been measured so far.

 For the lepton sector, the parallel results are
\begin{alignat}{1}
 &m_{e}=0.5116\:\mathrm{MeV}\,,\hspace{0.6cm}m_{\mu}=105.6\:\mathrm{MeV}\,,\hspace{1cm}m_{\tau}=1777\:\mathrm{MeV}\,;\nonumber\\
 & m_{1}=3.0\times10^{-4}\:\mathrm{eV}\,,\hspace{0.3cm} m_{2}=8.76\times10^{-3}\:\mathrm{eV}\,,\hspace{0.3cm}
   m_{3}=4.95\times10^{-2}\:\mathrm{eV}\,;\nonumber\\
 & s^{\,l}_{12}=0.560\,,\hspace{0.5cm} s^{\,l}_{23}=0.676\,,\hspace{0.5cm} s^{\,l}_{13}=0.0576\,,\nonumber\\
 & \delta^{\,l}=-0.0058\,\pi\,,\hspace{0.4cm} \beta_{1}=0.53\,\pi\,,\hspace{0.4cm} \beta_{2}=0.017\,\pi\,.
\end{alignat}
 The charged lepton masses are almost identical with those in the particle list \cite{1}. For the light left-handed neutrinos ,
 some quantities related directly to the experimental data are particularly calculated such as
\begin{alignat}{1}
 & \triangle m^{2}_{21}\approx 7.66\times10^{-5}\:\mathrm{eV^{2}}\,,\hspace{0.6cm}
   \triangle m^{2}_{32}\approx 2.37\times10^{-3}\:\mathrm{eV^{2}}\,,\nonumber\\
 & sin^{2}\theta^{\,l}_{12}\approx 0.314\,,\hspace{0.4cm} sin^{2}\theta^{\,l}_{23}\approx 0.457\,,\hspace{0.4cm}
   sin^{2}\theta^{\,l}_{13}\approx 0.0033\,,\nonumber\\
 & \langle m_{\beta\beta}\rangle\approx 2.7\times10^{-3}\:\mathrm{eV}\,,\hspace{0.8cm} J_{CP}^{\,l}\approx-2.4\times10^{-4}\,,
\end{alignat}
 where $\triangle m^{2}_{\alpha\beta}=m^{2}_{\alpha}-m^{2}_{\beta}$\,, and
\ba
 \langle m_{\beta\beta}\rangle=\left|m_{1}\left(c_{12}^{\,l}c_{13}^{\,l}e^{i\beta_{1}}\right)^{2}
                               +m_{2}\left(s_{12}^{\,l}c_{13}^{\,l}e^{i\beta_{2}}\right)^{2}
                               +m_{3}\left(s_{13}^{\,l}e^{-i\delta^{\l}}\right)^{2}\right|
\ea
 is the effective Majorana mass for neutrinoless double beta decay. These results are excellently in agreement with the recent neutrino oscillation experimental data \cite{20}. The heaviest one of the left-handed neutrino masses is less than $0.05$ eV. The value of $\theta^{\,l}_{13}$ is predicted to be $\thicksim3.3^{\circ}$. It is rather small but nonzero. In addition, there is also small $CP$-violating effect in the lepton sector because the three $CP$-violating phases are all non-vanishing values. The $J_{CP}^{\,l}$ value is predicted to be one magnitude higher than that in the quark sector. The value of $\langle m_{\beta\beta}\rangle$ is only of the order of $10^{-3}$, therefore it is very difficult to detect $0\nu\beta\beta$. Although measures about these quantities are still great challenges in the future neutrino experiments, we have confidence that all the predictions are promising to be tested in the near future.

 Finally, we also give the heavy right-handed neutrino masses (in GeV unit)
\ba
 M_{1}=3.5\times10^{8}\,,\hspace{0.5cm} M_{2}=1.7\times10^{9}\,,\hspace{0.5cm} M_{3}=1.8\times10^{11}\,.
\ea
 It is clear that their mass scale is nearly the middle point between the electroweak scale and the GUT scale. However, these right-handed Majorana neutrinos are too heavy to be found at the present colliders. Maybe they are possibly hunted in cosmic rays \cite{21}.

 To sum up the above numerical results, the model accurately fits the total twenty-five values about fermion masses and flavor mixings only by the six parameters. All the current measured values are exactly reproduced by our model, meanwhile, all the non-detected values are finely predicted in experimental limits. All the results are naturally produced without any fine tuning. That set of the mass matrices derived from the family symmetry and its breaking are key for the success of the model, among them, the parameter $\varepsilon_{0}$ plays a leading role. It is not only a source of the fermion mass hierarchy but also influences the flavor mixings, so it is actually a fundamental quantity in the model.

\vspace{1cm}
 \noindent\textbf{V. Conclusions}

\vspace{0.3cm}
 In the paper, we have suggested a new particle model based on the symmetry group
 $SU(3)_{C}\otimes SU(2)_{L}\otimes SU(2)_{L'}\otimes SU(2)_{R}\otimes U(1)_{B\mbox{-}L}\otimes SU(3)_{F}\otimes U(1)_{N}$\,.
 By means of the introduced super-heavy fermion, flavon and Higgs fields, the model carries out the two steps of breakings, namely the family symmetry breaking and the high-energy isospin symmetry breaking one after another. After the super-heavy fermions are all decoupling, we obtain the low-energy effective theory with the SM symmetry group. All the Yukawa coupling matrices show some regular structures and inherence relations. Among other things, the two key factors are the parameter $\varepsilon_{0}$ and the specific flavor structures which all stem from the family symmetry and it's breaking. After the electroweak breaking, all the fermion mass matrices are characterized only by the six parameters. The model can perfectly fit and explain all the current experimental data about the fermion masses and flavor mixings, in particular, it finely predicts the first generation quark masses and the values of $\theta^{\,l}_{13}$, $\langle m_{\beta\beta}\rangle$, $J_{CP}^{\,l}$ in neutrino physics. On all accounts, the model has only fewer parameters but shows a great prediction power, moreover, all the results are excellent and encouraging. This approach perhaps enlightens us on solving the flavor puzzles of the elemental fermions. Finally, we expect all the results to be tested in future experiments on the ground and in the sky. The experiments will undoubtedly provide us some important information about the flavor physics, and also help us to understand finely the mystery of the universe.

\vspace{1cm}
 \noindent\textbf{Acknowledgments}

\vspace{0.3cm}
 The author, W. M. Yang, would like to thank my mother and wife for long concern and love. My work is supported largely by them. This research is supported by chinese universities scientific fund.

\end{document}